\documentclass[conference]{IEEEtran}
\IEEEoverridecommandlockouts
\usepackage{cite}
\usepackage{amsmath,amssymb,amsfonts}
\usepackage{algorithmic}
\usepackage{graphicx}
\usepackage{textcomp}
\usepackage{xcolor}
\usepackage{url}
\usepackage{bm}
\usepackage{todonotes}
\usepackage{tikz}
\usetikzlibrary{decorations.pathreplacing} 
\usepackage{nicefrac}
\usepackage{subfigure}

\usetikzlibrary{arrows}
\tikzstyle{block} = [rectangle, draw, fill=white,
    text width=4em,
    text centered, rounded corners, minimum height=2em]
\tikzstyle{line} = [draw, -latex']
\definecolor{grayish}{gray}{0.4}
\definecolor{uibkblue}{cmyk}{1,0.6,0,0.65}%
\definecolor{uibkbluel}{rgb}{0.89,0.94,1.00}%

\definecolor{uibkorange}{cmyk}{0,.5,1,0}%
\definecolor{uibkorangel}{rgb}{1.00,0.90,0.76}%

\definecolor{uibkgray}{cmyk}{0,0,0,0.9}%
\definecolor{uibkgrayl}{cmyk}{0,0,0,0.6}%

\newcommand{\ie}{\textit{i}.\textit{e}.,}
\newcommand{\eg}{\textit{e}.\textit{g}.,}

\newcommand{\catamplitude}{\textsc{Amplitude}}
\newcommand{\catshape}{\textsc{Shape}}
\newcommand{\catgeometry}{\textsc{Geometry}}

\def\BibTeX{{\rm B\kern-.05em{\sc i\kern-.025em b}\kern-.08em
    T\kern-.1667em\lower.7ex\hbox{E}\kern-.125emX}}
\begin{document}

\title{
    \thanks{We thank Benedikt Lorch, Martin Bene\v{s}, Judith Senn, and Verena Lachner for valuable discussions and comments.
    Computational results were achieved using the LEO HPC infrastructure at the University of Innsbruck.}
    A Taxonomy of Miscompressions:\\Preparing Image Forensics for Neural Compression
    \vspace{-.75ex}
}
\author{\IEEEauthorblockN{Nora Hofer}
    \IEEEauthorblockA{
        \textit{University of Innsbruck, Austria}\\
        nora.hofer@uibk.ac.at}
    \and
    \IEEEauthorblockN{Rainer Böhme}
    \IEEEauthorblockA{
        \textit{University of Innsbruck, Austria}\\
        rainer.boehme@uibk.ac.at}
}

\makeatletter
\def\ps@IEEEtitlepagestyle{%
  \def\@oddfoot{\mycopyrightnotice}%
  \def\@oddhead{\hbox{}\@IEEEheaderstyle\leftmark\hfil\thepage}\relax
  \def\@evenhead{\@IEEEheaderstyle\thepage\hfil\leftmark\hbox{}}\relax
  \def\@evenfoot{}%
}
\def\mycopyrightnotice{%
  \begin{minipage}{\textwidth}
  \centering \scriptsize
  \copyright~2024 IEEE. Personal use of this material is permitted. Permission from IEEE must be obtained for all other uses, in any current or future media, including reprinting/republishing this material for advertising or promotional purposes, creating new collective works, for resale or redistribution to servers or lists, or reuse of any copyrighted component of this work in other works.
  \end{minipage}
}
\makeatother

\maketitle

\begin{abstract}
    Neural compression has the potential to revolutionize lossy image compression.
    Based on generative models, recent schemes achieve unprecedented compression rates at high perceptual quality, but they compromise semantic fidelity.
    Details of decompressed images may appear optically flawless, but semantically different from the originals, making compression errors difficult or impossible to detect.
    We explore the problem space and propose a provisional taxonomy of miscompressions.
    It defines three types of ``what happens'' and has a binary ``high impact'' flag indicating miscompressions that alter symbols.
    We discuss how the taxonomy can facilitate risk communication and research into mitigations.
\end{abstract}


\section{\label{sec:introduction}Introduction}

A turning point in the investigation of the 2013 Boston Marathon bombing was a bystander's cellphone photo that allowed police to identify one of the suspects in a crowd~\cite{ABCNews_2016}~\cite{justicegov_2024}.
Remarkably, the relevant part of the image comprised just 0.2\,\% of all pixels.
In this paper, we ask the question whether digital images will continue to serve as reliable sources in a future where neural compression becomes the default.

Neural image compression employs learning-based elements in the image compression pipeline, achieving high perceptual quality at unprecedented compression rates~\cite{yang2023introduction}.
State of the art schemes use generative networks to synthesize parts of an image~\cite{mentzer2020high,yang2024lossy}.
However, a drawback of this approach is that the synthesized details appear plausible and of high perceptual quality, but may be semantically different from the original.

\begin{figure}
    \centering
    \begin{center}
        \subfigure[Lossless original\newline ($4.1$ MB, $11.90$ bpp)]{\includegraphics[width=.155\textwidth]{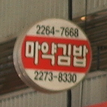}} 
        \subfigure[Neural compression\newline ($159$ kB, $0.46$ bpp, \cite{mentzer2020high})]{\includegraphics[width=.155\textwidth]{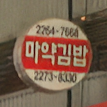}}
        \subfigure[JPEG QF 20\newline ($158$ kB, $0.46$ bpp) ]{\includegraphics[width=.155\textwidth]{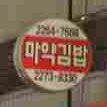}}
        \caption{
            State-of-the-art neural compression schemes can alter the semantic in details of the decompressed images.
            The high fidelity and the lack of visible compression artifacts make false reconstructions look more authentic than JPEG, which introduces visible distortion.
            (Crop of image $0831$ of DIV2K~\cite{Agustsson_2017DIV2K}, $0.41\%$ of the original.) All figures 
            are best viewed on screen and magnified.}
        \label{fig:intro}
    \end{center}
    \vspace{-3ex}
\end{figure}

To illustrate this, Fig.~\ref{fig:intro} compares a small crop (0.4\,\%) of an uncompressed image (left), to a version from the \textit{HiFiC} neural compression scheme~\cite{mentzer2020high} (middle), and the JPEG compressed image at quality factor (QF) 20 (right).
This factor was chosen to match the compression rate in bits per pixel between \textit{HiFiC} and JPEG.
While the neural compression retains clear readability of the numbers, closer inspection reveals that they differ:
the upper row changes from $2264-7668$ in the original to $22\textcolor{red}{{5}}4-766\textcolor{red}{{4}}$ in the \textit{HiFiC} reconstruction.
We confirmed that not only human observers, but also Google's Cloud AI optical character recognition came to this conclusion.
Given the high apparent quality of the image, observers unaware of its processing history might be inclined to fully trust the image and its semantic content.
By contrast, the visible compression artifacts in the JPEG image not only make the numbers difficult to read, but also signal low reliability and dissuade users from interpreting the numbers with confidence.

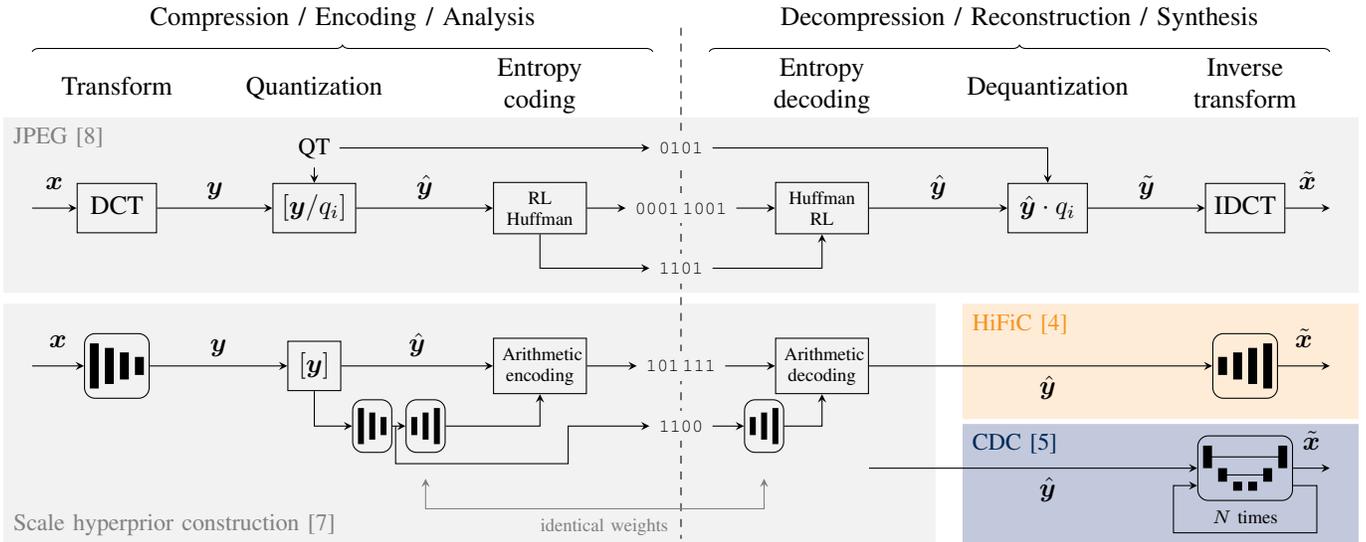
\begin{figure*}
    \begin{center}
        \begin{tikzpicture}[x=15mm,y=16mm,>=stealth]

    \draw [decorate,decoration={brace}] (-5.75,.3)-- node [above=1ex] {Compression / Encoding / Analysis} (-.25,.3);
    \draw [decorate,decoration={brace}] (.25,.3)-- node [above=1ex] {Decompression / Reconstruction / Synthesis} (5.75,.3);

    \draw (-5,0) node  (T) {Transform\strut};
    \draw (-3.25,0) node  (Q) {Quantization\strut};
    \draw (-1.25,0) node  (E) {\parbox{2cm}{\centering Entropy coding\strut}};

    \draw (5,0) node  (IT) {\parbox{2cm}{\centering Inverse transform\strut}};
    \draw (3.25,0) node  (IQ) {Dequantization\strut};
    \draw (1.25,0) node  (IE) {\parbox{2cm}{\centering Entropy decoding\strut}};


    \begin{scope}[overlay]
        \draw [uibkorange!15,fill] (6,-2.75) rectangle (2.5,-1.8) node [below right,uibkorange] {\small HiFiC~\cite{mentzer2020high}};
        \draw [uibkblue!15,fill] (6,-3.8) rectangle (2.5,-2.8) node [below right,uibkblue] {\small CDC~\cite{yang2024lossy}};
        \draw [black!5,fill] (2.25,-1.8) rectangle (-6,-3.8) node [above right,gray] {\small Scale hyperprior construction~\cite{balle2018variational}} ;
        \draw [black!5,fill] (6,-1.7) rectangle (-6,-.25) node [below right,gray] {\small JPEG~\cite{wallace1991jpeg}} ;
    \end{scope}

    \draw [dashed] (0,.5)--++(0,-4.25);


    \draw (T|-0,-1) node [draw,minimum width=3em] (dct) {$\text{DCT}$\strut};
    \draw (IT|-0,-1) node [draw,minimum width=3em] (idct) {$\text{IDCT}$\strut};

    \draw (Q|-0,-1) node [draw,minimum width=3em] (qt) {$[\bm{y}/q_i]$\strut};
    \draw (IQ|-0,-1) node [draw,minimum width=3em] (iqt) {$\hat{\bm{y}} \cdot q_i$\strut};

    \draw (E|-0,-1) node [draw,minimum width=3em] (huff) {\scriptsize\parbox{1cm}{\centering RL\\Huffman}};
    \draw (IE|-0,-1) node [draw,minimum width=3em] (ihuff) {\scriptsize\parbox{1cm}{\centering Huffman\\RL}};

    \draw (qt|-0,-.5) node (qtt) {\small QT};
    \draw [->] (qtt)--(qt);

    \draw (0,-1) node [fill=black!5] (jbits) {\scriptsize\texttt{0001\,1001}};
    \draw (0,-1|-qtt) node [fill=black!5] (qtbits) {\scriptsize\texttt{0101}};
    \draw (0,-1.5) node [fill=black!5] (hbits) {\scriptsize\texttt{1101}};

    \draw [->] (huff) |- (hbits);
    \draw [->] (hbits) -| (ihuff);

    \draw [->] (huff)--(jbits);
    \draw [->] (jbits)--(ihuff);

    \draw [->] (qtt)--(qtbits);
    \draw [->] (qtbits)-|(iqt);

    \draw [->] (dct)-- node [above] {$\bm{y}$} (qt);
    \draw [->] (qt)-- node [above] {$\hat{\bm{y}}$} (huff);
    \draw [->] (ihuff)-- node [above] {$\hat{\bm{y}}$} (iqt);
    \draw [->] (iqt)-- node [above] {$\tilde{\bm{y}}$} (idct);

    \draw [<-] (dct)-- node [above] {$\bm x$\strut} ++(-.75,0);
    \draw [->] (idct)-- node [above] {$\tilde{\bm x}\strut$} ++(.75,0);


    \begin{scope}[yshift=-5mm]

        \draw (T|-0,-2) node [draw,inner sep=2pt,rounded corners] (enc) {\tikz[line width=.75ex,x=2mm,y=.6mm]{
                \draw (0,-5)--++(0,10);
                \draw (1,-4)--++(0,8);
                \draw (2,-3)--++(0,6);
                \draw (3,-2)--++(0,4);
            }};

        \draw (-2.5,-2.5) node [draw,inner sep=2pt,left=.25em,rounded corners] (ha) {\tikz[line width=.5ex,x=1.5mm,y=.6mm]{
                \draw (1,-4)--++(0,8);
                \draw (2,-3)--++(0,6);
                \draw (3,-2)--++(0,4);
            }};

        \draw (-2.5,-2.5) node [draw,inner sep=2pt,right=.25em,rounded corners] (hs) {\tikz[line width=.5ex,x=1.5mm,y=.6mm]{
                \draw (-1,-4)--++(0,8);
                \draw (-2,-3)--++(0,6);
                \draw (-3,-2)--++(0,4);
            }};

        \draw (.5,-2.5) node [draw,inner sep=2pt,right=.25em,rounded corners] (hs2) {\tikz[line width=.5ex,x=1.5mm,y=.6mm]{
                \draw (-1,-4)--++(0,8);
                \draw (-2,-3)--++(0,6);
                \draw (-3,-2)--++(0,4);
            }};

        \draw (IT|-0,-2) node [draw,inner sep=2pt,rounded corners] (gangen) {\tikz[line width=.75ex,x=2mm,y=.6mm]{
                \draw (0,-5)--++(0,10);
                \draw (-1,-4)--++(0,8);
                \draw (-2,-3)--++(0,6);
                \draw (-3,-2)--++(0,4);
            }};

        \draw (IT|-0,-2.85) node [draw,inner sep=2pt,rounded corners] (diffgen) {\tikz[line width=.75ex,x=2mm,y=.6mm]{
                \draw (0,5)--++(0,5);
                \draw (1,2)--++(0,3);
                \draw (2,0)--++(0,2);
                \draw (3,0)--++(0,2);
                \draw (4,2)--++(0,3);
                \draw (5,5)--++(0,5);
                \draw [thin] (0,7.5)--++(5,0);
                \draw [thin] (1,3.5)--++(3,0);
            }};

        \draw (Q|-0,-2) node [draw,minimum width=2em] (nqt) {$[\bm{y}]$\strut};
        \draw (E|-0,-2) node [draw,minimum width=3em] (ae) {\scriptsize\parbox{1cm}{\centering Arithmetic\\encoding}};
        \draw (IE|-0,-2) node [draw,minimum width=3em] (iae) {\scriptsize\parbox{1cm}{\centering Arithmetic\\decoding}};

        \draw (0,-2) node [fill=black!5] (lbits) {\scriptsize\texttt{101\,111}};
        \draw (0,-2.5) node [fill=black!5] (pbits) {\scriptsize\texttt{1100}};

    \end{scope}
    \draw [->] (enc)-- node [above] {$\bm{y}$} (nqt);
    \draw [->] (nqt)-- node [above] {$\hat{\bm{y}}$} (ae);
    \draw [->] (iae)-- (gangen);
    \draw (IQ|-gangen) node [below] {$\hat{\bm{y}}$};
    \draw [->] (iae.east|-diffgen)-- (diffgen);
    \draw (IQ|-diffgen) node [below] {$\hat{\bm{y}}$};
    \draw [<-] (enc)-- node [above] {$\bm x$\strut} ++(-.75,0);
    \draw [->] (gangen)-- node [above] {$\tilde{\bm x}\strut$} ++(.75,0);
    \draw [->] (diffgen)-- node [above] {$\tilde{\bm x}\strut$} ++(.75,0);
    \draw [->] (nqt) |- (ha);
    \draw [->] (ha) -- coordinate [pos=.25] (qq) (hs);
    \draw [->] (pbits) -- (hs2);
    \draw [->] (hs) -| (ae);
    \draw [->] (hs2) -| (iae);
    \draw [->] (ae) -- (lbits);
    \draw [->] (lbits) -- (iae);
    \draw [->] (qq) |- (-1,-3.125) |- (pbits);
    \draw (diffgen.-20) -- ++(2ex,0) |- (diffgen|-0,-3.7);
    \draw [<-] (diffgen.-160) -- ++(-2ex,0) |- (diffgen|-0,-3.7) node [above] {\scriptsize $N$ times};
    \draw [<->,gray] (hs|-0,-3.25) |-  node [pos=.85,below] {\scriptsize identical weights} (0,-3.5) -| (hs2|-0,-3.25);
\end{tikzpicture}
    \end{center}
    \caption{Comparison of the conventional JPEG compression (top) and the neural compression schemes used in this paper (bottom). Elements with rounded corners are CNNs trained on datasets and used in inference mode during encoding and reconstruction.}
    \label{fig:pipeline}
    \vspace{-3ex}
\end{figure*}

In this paper, we propose the term ``miscompression'' to describe semantic changes resulting from lossy compression.
This new%
\footnote{The closest related work we are aware of is an attempt to introduce copy-evident marks into images which only appear after JPEG compression~\cite{lewis2009towards}.} 
phenomenon arose with neural compression and deserves the attention of researchers and forensic practitioners.
To facilitate the conversation, we develop a taxonomy of miscompressions based on the explorative visual inspection of three benchmark datasets, examining five different neural compression schemes at different quality settings.
The examples used for illustration in this paper are produced using two schemes of the latest generation.
We derive implications for forensics and society at large, and outline objectives for research into mitigations.

We emphasize that all authors of recent neural compression schemes generally acknowledge the risk of hallucinations, warn against using their schemes for critical applications~\cite[p.~10]{mentzer2020high}, and occasionally point to examples of text becoming unreadable.
However, miscompressions take many forms and pose a risk to society if left unaddressed.
We see our work as a first step towards solutions.

The paper is structured as follows.
Section~\ref{sec:background} introduces the principles of neural compression by comparing it to the JPEG pipeline.
Section~\ref{sec:miscompressions} defines miscompressions, describes our empirical approach, and presents the proposed taxonomy using examples.
Section~\ref{sec:discussion} discusses implications and outlines a research agenda.
Section~\ref{sec:conclusion} concludes our paper.

\section{\label{sec:background}Primer on neural compression}

The top row of Fig.~\ref{fig:pipeline} shows the key components of the lossy image compression and decompression pipeline.
The first component \textbf{transforms} an input image from the spatial domain into a domain where pixels are decorrelated and the variance is concentrated in fewer coefficients.
A useful property of common transformations is that the distribution of the coefficients can be approximated parametrically.
The second component, \textbf{quantization}, is a deliberately lossy process.
It maps a range of values to a single discrete value.
The quantization steps can be adjusted according to the relevance of each coefficient for the reconstruction of the signal.
The final component of the compression pipeline is lossless \textbf{entropy coding}, which approximates Shannon's theorem~\cite{shannon1948mathematical} by encoding a sequence of quantized values into a sequence of bits.
The better the input distribution is known, the tighter the approximation and the shorter the resulting bitstream.

Over decades, JPEG~\cite{wallace1991jpeg} has been the most popular lossy image compression scheme.
It implements the transform by  the linear discrete cosine transform (DCT)~\cite{ahmed1974DCT} on non-overlapping blocks of $8\times 8$ pixels.
The resulting coefficients are then quantized by dividing them by frequency-specific quantization factors and subsequent rounding to the nearest integer.
The quantized coefficients are arranged in zigzag order and entropy-coded using run-length (RL) and Huffman encoding~\cite{huffman1952method}.
The resulting image file contains the quantization tables (QT), the quantized DCT coefficients, and the Huffman tables.
JPEG decompression reverts this sequence of steps.

Neural compression replaces components of this pipeline with learnable elements, typically deep convolutional neural networks (CNN).
This emerging field has its own jargon.
Encoding and analysis are used to describe compression.
Reconstruction and synthesis denote decompression (cf. Fig.~\ref{fig:pipeline}).

Learning the transform promises that irrelevance in the input signal can be isolated better in the so-called latent space than with known structured transformations, such as block-wise DCT.
While the networks have shown to derive basis functions similar to those in linear transforms~\cite{duan2022opening}, nonlinear transforms offer better adaptation to varying data distributions and can be optimized for specific distortion metrics.
The loss function used for training has two terms: distortion and rate.
By weighting these terms, different tradeoffs between image quality and file size can be achieved.
Finding the right distortion metric for neural compression is an active field of research~\cite{ding2020image, NEURIPS2020_00482b9b}.
Once trained, the weights are stored in the encoder and decoder.
The CNNs are used in inference mode for the encoding and reconstruction of images.

Quantization in neural compression typically involves rounding and truncation~\cite{theis2017lossy,balle2016end}.
Unlike JPEG, it does not use and transmit QTs.
The quantization step size is controlled by the scaling in last layer of the transform CNN and thus learned.
Therefore, neural compression schemes commonly require a separately trained model for each target quality.

Also, entropy coding requires modifications.
A drawback of learning a transform is the lack of a statistical model of the latent space.
Here, the answer to machine learning is machine learning.
The distribution of the latent space is modelled with a trained auto encoder.
The prediction of this model is used to parameterize an arithmetic encoder.
As the distribution is data dependent, the latent space of this prediction model must itself be transmitted to the decoder to enable reconstruction.
Ballé et al.'s scale hyperprior construction~\cite{balle2018variational} is the basis of the two schemes evaluated in this work.

\textit{HiFiC}~\cite{mentzer2020high} and \textit{CDC}~\cite{yang2024lossy} improve on previous approaches to neural compression by using generative models for the inverse transform.
\textit{HiFiC} uses a \textbf{generative adversarial network} (GAN).
GANs are trained with a rivaling discriminator network that regularizes the generator network towards producing outputs of high perceptual quality~\cite{goodfellow2014generative}.
To control the content of the reconstruction, the generator is conditioned with the latent representation of the encoded image.
\textit{CDC} implements a \textbf{diffusion model}~\cite{sohl2015deep} for the inverse transform.
It uses the latent representation to condition the denoising diffusion probabilistic model~\cite{ho2020denoising} (DDPM).
Both schemes are trained end-to-end, allowing the variational autoencoder in the transform component to learn how to turn an input image to a condition.
The rate estimate of the loss function is taken from the hyperprior model and the distortion estimate is a weighted sum of the perceptual loss and the mean squared error between the input and the reconstructed image.
Increasing the weight of the perceptual metric gives the model the flexibility to deviate from the input signal and ``make up'' details during reconstruction.
This enables high perceptual quality at unprecedented compression rates, but compromises semantic fidelity.

The digital image forensics community has only recently started to address the impact of neural compression:
Berthet et al.\ revisit copy-move forgery detection~\cite{berthet2022ai} and source social network identification~\cite{berthet2023impact},
Bergmann et al.\ use traces in the frequency and spatial domain for detection~\cite{bergmann2023frequency,bergmann2024forensic}, and
Chen et al.\ show a vulnerability of different neural compression schemes to adversarial perturbations in the input image~\cite{chen2023towards}.
Jalilian et al.\ propose CNNs to compress biometric images~\cite{jalilian2022}. 
However, to the best of our knowledge, nobody has yet investigated semantic changes and their implications.

\section{\label{sec:miscompressions}Miscompressions}

Semantic interpretation is the understanding of a perceived scene by applying domain terminology, \ie{} semantic concepts~\cite{marr2010vision}.
It is carried out by a human observer and is heavily influenced by their prior semantic and conceptual knowledge of the domain~\cite{hudelot2005symbol}.
The \emph{semantic meaning} of a scene or an image is the result of semantic interpretation.

We define \textbf{miscompressions} as reconstruction errors that occur when there is a discrepancy between the semantic meaning of an original image (detail) and its reconstructed version after neural compression.
As a test, we require that a human observer asked to verbally describe the relevant part of the image would come up with a different description.
Note that this definition applies to entire images as well as individual image details.
Digital images used as evidence in forensic investigations often capture relevant details unconsciously, as can be seen from the Boston Marathon bombing example.
Such photographs often preserve details that serve as objective representations of reality.
Consequently, forensic investigators typically focus on analyzing the semantic meaning of specific image details, such as objects or individuals in the background, rather than interpreting the semantics of the entire image.

Miscompressions are a new phenomenon, requiring a precise terminology to describe, mitigate, or ultimately avoid them.
This paper takes a first step in this direction and proposes a provisional taxonomy, systematically derived with three key objectives in mind.
First, the taxonomy should facilitate research into the risks posed by miscompressions.
Distinguishing between different forms of reconstruction errors that lead to miscompressions allows us to measure their prevalence, and compare compression schemes and the optimization metrics used.
Second, it should facilitate research towards making neural compression safer.
While ideal compression schemes would be completely immune to miscompressions, it is uncertain whether this is achievable at competitive compression rates.
However, confidence in neural compression would be greatly improved if it could be ensured that certain types of miscompressions are extremely unlikely.
The third intended application of our taxonomy is to deal with the remaining risk in practice.
It should allow forensics experts to explain miscompressions using references to scientific evidence in order to convince a judge or jury.

\begin{figure}
    \begin{tikzpicture}
        \begin{scope}
            \draw (0,0) node [left=.25em,inner sep=0] (A)  {\includegraphics[width=3cm]{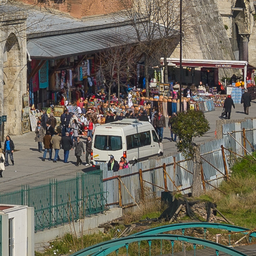}};
            \clip (A.south east) rectangle (A.north west);
            \draw (A)++(.5,.5) node {\includegraphics[width=12cm]{fig/crop_256_div0890_original}};
            \draw (A.north) node [below,white] {Original};
        \end{scope}

        \begin{scope}
            \draw (0,0) node [right=.25em,inner sep=0] (B)  {\includegraphics[width=3cm]{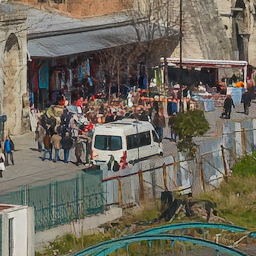}};
            \clip (B.south east) rectangle (B.north west);
            \draw (B)++(.5,.5) node {\includegraphics[width=12cm]{fig/crop_256_div0890_hifichi_semanticchanged_breaklightsoff}};
            \draw (B.north) node [below,white] {Reconstructed};
        \end{scope}

        \draw (A.north-|-5.5,0) node [below right] (O) {\textit{\textbf{HiFiC Hi}}};

        \draw (O.west|-0,0)++(0,4ex) node [right] {\scriptsize{compressed:}};
        \draw (O.west|-0,0)++(0,2ex) node [right] {\scriptsize0.48 bpp};
        \draw (O.west|-0,0)++(0,0ex) node [right] {\scriptsize2048$\times$1368};

        \draw (O.west|-0,0)++(0,-4ex) node [right] {\scriptsize{cropped scene:}};
        \draw (O.west|-0,0)++(0,-6ex) node [right] {\scriptsize{64$\times$64 {($0.15\%$)}}};
    \end{tikzpicture}
    \medskip

    \begin{tikzpicture}
        \begin{scope}
            \clip (A.south east) rectangle (A.north west);
            \draw (A)++(-.2,-.25) node {\includegraphics[width=12cm]{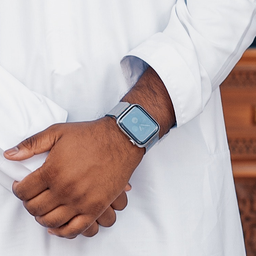}};
            \draw (A.north) node [below,white] {Original};
        \end{scope}

        \begin{scope}
            \draw (0,0) node [right=.25em,inner sep=0] (B)  {\includegraphics[width=3cm]{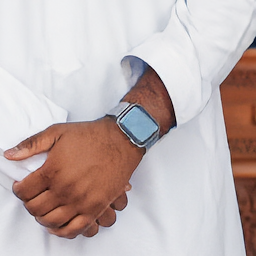}};
            \clip (B.south east) rectangle (B.north west);
            \draw (B)++(-.2,-.25) node {\includegraphics[width=12cm]{fig/uhr_hificmi.png}};
            \draw (B.north) node [below,white] {Reconstructed};
        \end{scope}

        \draw (A.north-|-5.5,0) node [below right] (O) {\textit{\textbf{HiFiC Mi}}};

        \draw (O.west|-0,0)++(0,4ex) node [right] {\scriptsize{compressed:}};
        \draw (O.west|-0,0)++(0,2ex) node [right] {\scriptsize0.19 bpp}; 
        \draw (O.west|-0,0)++(0,0ex) node [right] {\scriptsize1365$\times$2048};

        \draw (O.west|-0,0)++(0,-4ex) node [right] {\scriptsize{cropped scene:}};
        \draw (O.west|-0,0)++(0,-6ex) node [right] {\scriptsize{64$\times$64 {($0.15\%$)}}};

    \end{tikzpicture}

    \medskip

    \begin{tikzpicture}
        \begin{scope}
            \draw (0,0) node [left=.25em,inner sep=0] (A)  {\includegraphics[width=3cm]{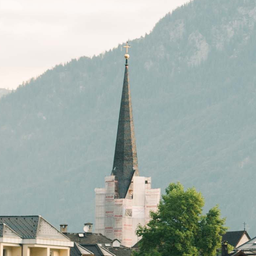}};
            \clip (A.south east) rectangle (A.north west);
            \draw (A)++(0,-1.5) node {\includegraphics[width=6cm]{fig/crop7a663d483f843a589bc41698ce3257e7.png}};
            \draw (A.north) node [below,black] {Original};
        \end{scope}

        \begin{scope}
            \draw (0,0) node [right=.25em,inner sep=0] (B)  {\includegraphics[width=3cm]{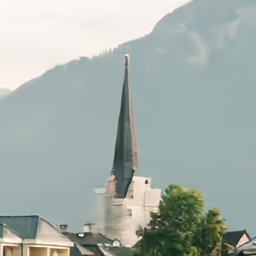}};
            \clip (B.south east) rectangle (B.north west);
            \draw (B)++(0,-1.5) node {\includegraphics[width=6cm]{fig/crop_7a663d483f843a589bc41698ce3257e7_param00.png}};
            \draw (B.north) node [below,black] {Reconstructed};
        \end{scope}

        \draw (A.north-|-5.5,0) node [below right] (O) {\textit{\textbf{CDC 0.0}}};

        \draw (O.west|-0,0)++(0,4ex) node [right] {\scriptsize{compressed:}};
        \draw (O.west|-0,0)++(0,2ex) node [right] {\scriptsize0.08 bpp}; 
        \draw (O.west|-0,0)++(0,0ex) node [right] {\scriptsize1152$\times$1920};

        \draw (O.west|-0,0)++(0,-4ex) node [right] {\scriptsize{cropped scene:}};
        \draw (O.west|-0,0)++(0,-6ex) node [right] {\scriptsize{128$\times$128 {($0.74\%$)}}};

    \end{tikzpicture}
    \medskip


    \begin{tikzpicture}
        \begin{scope}
            \draw (0,0) node [left=.25em,inner sep=0] (A)  {\includegraphics[width=3cm]{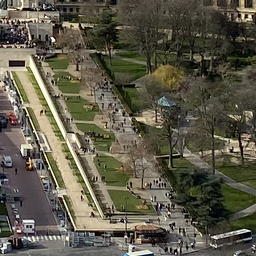}};
            \clip (A.south east) rectangle (A.north west);
            \draw (A)++(1,-.25) node {\includegraphics[width=6cm]{fig/paris_1ac06ad4b53f9880698aaaaa730363c6.png}};
            \draw (A.north) node [below,white] {Original};
        \end{scope}

        \begin{scope}
            \draw (0,0) node [right=.25em,inner sep=0] (B)  {\includegraphics[width=3cm]{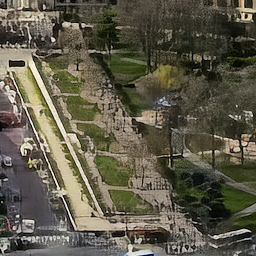}};
            \clip (B.south east) rectangle (B.north west);
            \draw (B)++(1,-.25) node {\includegraphics[width=6cm]{fig/paris_1ac06ad4b53f9880698aaaaa730363c6_hificlo.png}};
            \draw (B.north) node [below,white] {Reconstructed};
        \end{scope}

        \draw (A.north-|-5.5,0) node [below right] (O) {\textit{\textbf{HiFiC Lo}}};

        \draw (O.west|-0,0)++(0,4ex) node [right] {\scriptsize{compressed:}};
        \draw (O.west|-0,0)++(0,2ex) node [right] {\scriptsize0.17 bpp}; 
        \draw (O.west|-0,0)++(0,0ex) node [right] {\scriptsize2016$\times$1512};

        \draw (O.west|-0,0)++(0,-4ex) node [right] {\scriptsize{cropped scene:}};
        \draw (O.west|-0,0)++(0,-6ex) node [right] {\scriptsize{128$\times$128 {($0.54\%$)}}};

    \end{tikzpicture}
    \caption{Category \catamplitude{}: Reconstructions differ in the amplitude of spatial frequencies in the signal, affecting attributes such as brightness, color saturation, and the intensity of high frequency components.}
    \label{fig:ex-amplitude}
    \vspace{-2ex}
\end{figure}

\subsection{Method}
Our approach is exploratory.
We focus on five relevant neural compression schemes~\cite{balle2016end,balle2018variational,minnen2020channel,mentzer2020high,yang2024lossy}, compress the test images of three widely-used benchmark datasets, \textit{CLIC2020}~\cite{CLIC2020}, \textit{DIV2K}~\cite{Agustsson_2017DIV2K}, and \textit{Kodak}~\cite{franzen1999kodak} (full dataset), and manually inspect the reconstructions of a total of $552$ images to identify miscompressions, using difference images for guidance, where necessary.

We have observed miscompressions in all tested schemes, but decided to shift our focus to \textit{HiFiC}~\cite{mentzer2020high} and \textit{CDC}~\cite{yang2024lossy}.
These schemes employ generative networks and stand out, as their reconstructions are of such high fidelity that they appear deceptively authentic.
We use the pre-trained \textit{HiFiC} model with 180 million parameters for the three available compression intensities \emph{high}, \emph{mid}, and \emph{low},%
\footnote{\textit{HiFiC}: \url{https://github.com/tensorflow/compression/tree/master/models/hific}}
and the pre-trained $\mathcal{X}$-parameterization model of \textit{CDC} with 54 million parameters for the widely-adopted LPIPS loss at weights $0.0$ and $0.9$.%
\footnote{\textit{CDC}: \url{https://github.com/buggyyang/CDC_compression}}
We varied the noise seeds for \textit{CDC} and found that miscompressions prevailed.

\subsection{Taxonomy}
At a high level, our taxonomy separates the signal processing perspective (``What happens?'') from the semantic impact (``How bad is the misinterpretation?'').
Based on the apparent transformation of the signal, we define three categories.
To illustrate each category we provide examples, cropped from compressed images of three datasets, and
specify the compression model used, the bpp of the compressed image, the pixel dimensions of the original image, and the crop, as well as the percentage of the original represented by the displayed crop.
The selected examples illustrate the characteristics of each type of miscompression.
In practice, many miscompressions exhibit combined effects of multiple types.

\paragraph{\catamplitude}
refers to reconstructions that differ in the amplitude of spatial frequencies in the image signal, such as changes in brightness, color saturation, or intensity of high frequency components.
Attenuation seems to be more common, although we cannot rule out amplification.
Unlike global signal processing operations, these effects tend to be local and content-dependent.
Objects that we have found to be particularly susceptible to this type of miscompression include lights, colors of eye, hair, and skin, as well as birthmarks, and tattoos.
Attenuation can result in altered colors or ``disappearing'' objects.
Semantic changes occur when the amplitude carries meaning, as illustrated in the examples in Fig.~\ref{fig:ex-amplitude}.
In the top row, the fact that the car was braking, as indicated by the brake lights, is lost in the reconstruction.
In the second row, a reconstructed watch appears to be turned off but is actually on and displaying the time.
In the third row, the reconstructed image of a church tower does no longer include the Christian cross.
In the bottom row, the reconstructed image of a park does not include the people present in the original image.

\begin{figure}
    \begin{tikzpicture}[remember picture]
        \draw (0,0) node [inner sep=0] (A) {\includegraphics[width=.975\linewidth]{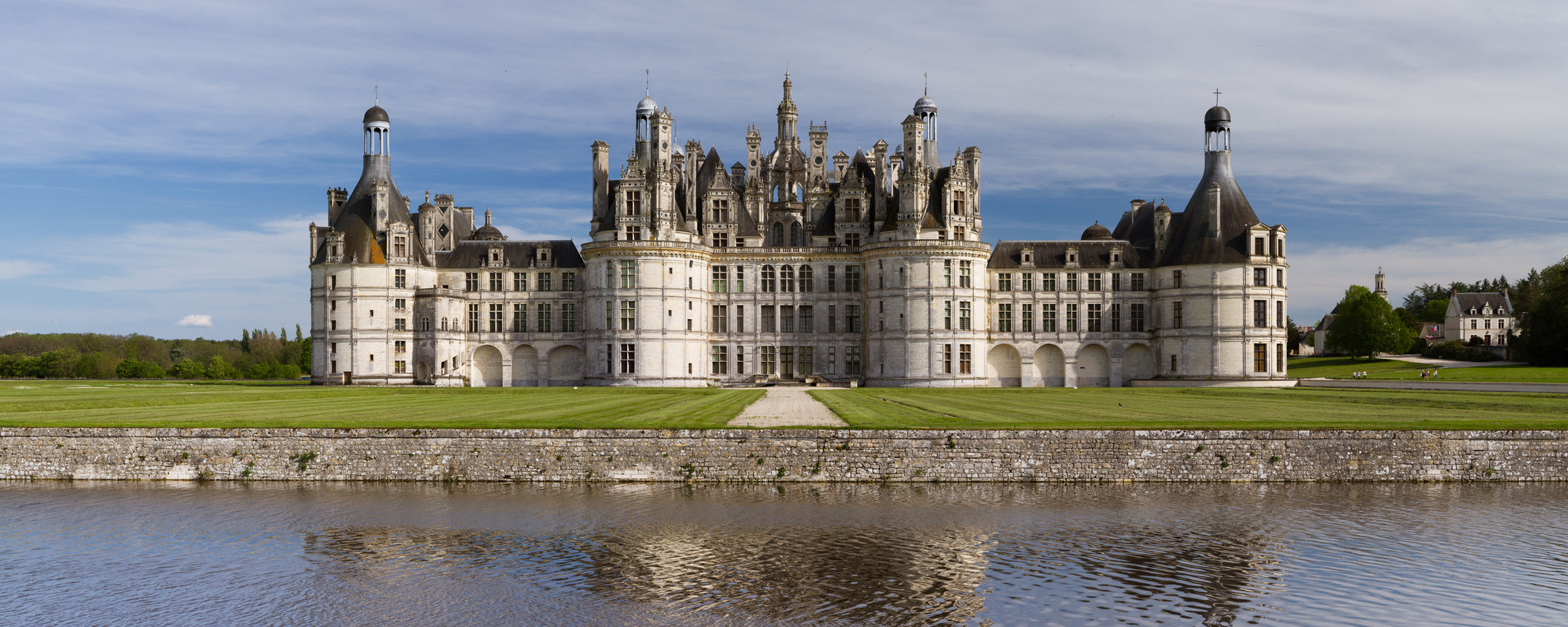}};
        \draw (A.east)++(-20.5pt,-2pt) node [draw,inner sep=0,white,minimum width=9pt,minimum height=9pt] (Q) {~};
    \end{tikzpicture}

    \medskip

    \begin{tikzpicture}[remember picture]
        \begin{scope}
            \draw (0,0) node [left=.25em,inner sep=0] (A)  {\includegraphics[width=3cm]{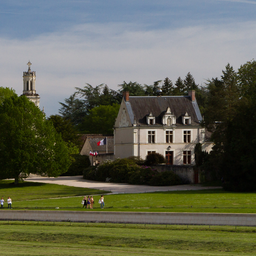}};
            \clip (A.south east) rectangle (A.north west);
            \draw (A)++(1,.4) node {\includegraphics[width=12cm]{fig/crop_265_div0830_original}};
            \draw (A.north) node [below,white] {Original};
        \end{scope}

        \begin{scope}
            \draw (0,0) node [right=.25em,inner sep=0] (B)  {\includegraphics[width=3cm]{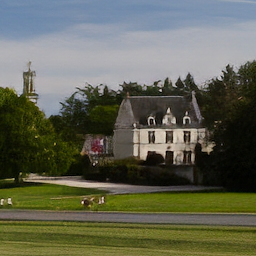}};
            \clip (B.south east) rectangle (B.north west);
            \draw (B)++(1,.4) node {\includegraphics[width=12cm]{fig/crop_265_div0830_hificlo_semanticchanged_grafitti}};
            \draw (B.north) node [below,white] {Reconstructed};
        \end{scope}

        \draw [overlay,white] (A.north west)--(Q.north west);
        \draw [overlay,black,dashed] (A.north west)--(Q.north west);
        \draw [overlay,white] (A.south east)--(Q.south east);
        \draw [overlay,black,dashed] (A.south east)--(Q.south east);

        \draw (A.north-|-5.5,0) node [below right] (O) {\textit{\textbf{HiFiC Lo}}};

        \draw (O.west|-0,0)++(0,4ex) node [right] {\scriptsize{compressed:}};
        \draw (O.west|-0,0)++(0,2ex) node [right] {\scriptsize0.18 bpp};
        \draw (O.west|-0,0)++(0,0ex) node [right] {\scriptsize2040$\times$816};

        \draw (O.west|-0,0)++(0,-4ex) node [right] {\scriptsize{cropped scene:}};
        \draw (O.west|-0,0)++(0,-6ex) node [right] {\scriptsize{64$\times$64 {($0.25\%$)}}};

    \end{tikzpicture}

    \medskip


    \begin{tikzpicture}
        \begin{scope}
            \draw (0,0) node [left=.25em,inner sep=0] (A)  {\includegraphics[width=3cm]{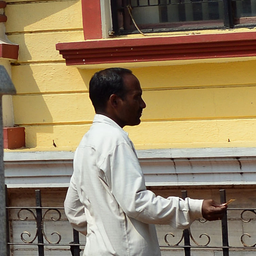}};
            \clip (A.south east) rectangle (A.north west);
            \draw (A)++(.2,-1.4) node {\includegraphics[width=10cm]{fig/crop_0891.png}};
            \draw [black,very thin,step=2em,xshift=0em] (A.north east) grid (A.south west);
            \draw (A.north) node [below,white] {Original};
        \end{scope}

        \begin{scope}
            \draw (0,0) node [right=.25em,inner sep=0] (B)  {\includegraphics[width=3cm]{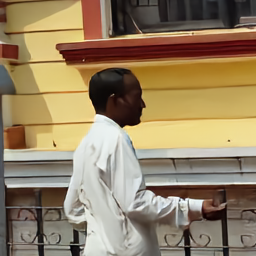}};
            \clip (B.south east) rectangle (B.north west);
            \draw (B)++(.2,-1.4) node {\includegraphics[width=10cm]{fig/0891_xparam00}};
            \draw [black,very thin,step=2em,xshift=1em] (B.north east) grid (B.south west);
            \draw (B.north) node [below,white] {Reconstructed};
        \end{scope}

        \draw (A.north-|-5.5,0) node [below right] (O) {\textit{\textbf{CDC 0.0}}};

        \draw (O.west|-0,0)++(0,4ex) node [right] {\scriptsize{compressed:}};
        \draw (O.west|-0,0)++(0,2ex) node [right] {\scriptsize0.24 bpp}; 
        \draw (O.west|-0,0)++(0,0ex) node [right] {\scriptsize1536$\times$1920};

        \draw (O.west|-0,0)++(0,-4ex) node [right] {\scriptsize{cropped scene:}};
        \draw (O.west|-0,0)++(0,-6ex) node [right] {\scriptsize{64$\times$64 {($0.14\%$)}}};
    \end{tikzpicture}
    \medskip

    \begin{tikzpicture}
        \begin{scope}
            \draw (0,0) node [left=.25em,inner sep=0] (A)  {\includegraphics[width=3cm]{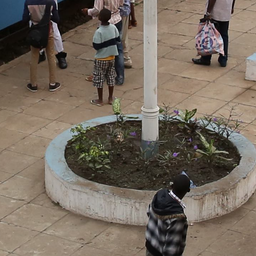}};
            \clip (A.south east) rectangle (A.north west);
            \draw (A)++(.5,-1) node {\includegraphics[width=12cm]{fig/div0850_original}};
            \draw (A.north) node [below,white] {Original};
        \end{scope}

        \begin{scope}
            \draw (0,0) node [right=.25em,inner sep=0] (B)  {\includegraphics[width=3cm]{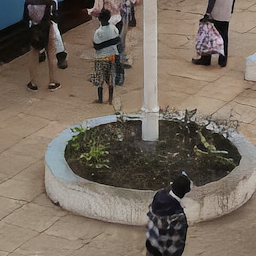}};
            \clip (B.south east) rectangle (B.north west);
            \draw (B)++(.5,-1) node {\includegraphics[width=12cm]{fig/example_leaf_div0850_hificlo_semanticchanged_color_shirt}};
            \draw (B.north) node [below,white] {Reconstructed};
        \end{scope}

        \draw (A.north-|-5.5,0) node [below right] (O) {\textit{\textbf{HiFiC Lo}}};

        \draw (O.west|-0,0)++(0,4ex) node [right] {\scriptsize{compressed:}};
        \draw (O.west|-0,0)++(0,2ex) node [right] {\scriptsize0.17 bpp}; 
        \draw (O.west|-0,0)++(0,0ex) node [right] {\scriptsize2040$\times$1365};

        \draw (O.west|-0,0)++(0,-4ex) node [right] {\scriptsize{cropped scene:}};
        \draw (O.west|-0,0)++(0,-6ex) node [right] {\scriptsize{64$\times$64 {($0.15\%$)}}};

    \end{tikzpicture}

    \caption{Category \catgeometry{}: Reconstructions contain geometric transformations, such as translation, rotation, scaling, and shearing, including shifted shapes and dissolved contours. (The top image illustrates the level of detail at which miscompressions occur. A grid was added to the middle crop.)}
    \label{fig:ex-geometry}
    \vspace{-3ex}
\end{figure}

\paragraph{\catgeometry} refers to reconstructions with geometric transformations such as translations, rotation, scaling, and shearing.
This includes locally shifted shapes, dissolved contours, and imperfect representations of 3D scenes in 2D pixel matrices.
Semantic changes occur when the geometry of an object carries semantic meaning, as illustrated in Fig.~\ref{fig:ex-geometry}.
The top row shows a reconstruction of a French flag that could be mistaken for graffiti on the wall in background.
The direction of the nose and chin of the person's profile in the second row is altered and differs from the original.
The bottom row shows a leaf in the foreground that looks like a crack in the floor after reconstruction.
Other susceptible objects, not illustrated here, include shadows and reflections.
Semantic changes occur if the direction of a shadow changes and suggests a different positioning of the object casting the shadow,
or if an object that is placed in front of a reflecting body of water or glass appears to be part of the reflection.
This has implications for forensic methods that exploit inconsistencies in lighting direction and shadows for the detection of image manipulations~\cite[p.~14]{piva2013overview}.

\begin{figure}
    \begin{tikzpicture}

        \begin{scope}
            \draw (0,0) node [left=.25em,inner sep=0] (A)  {\includegraphics[width=3cm]{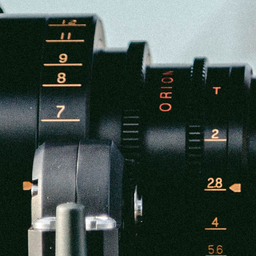}};
            \clip (A.south east) rectangle (A.north west);
            \draw (A)++(3,-2) node {\includegraphics[width=12cm]{fig/crop_clic400984b87394ada6d9627ed918908986_original}};
            \draw (A.north) node [below,white] {Original};
        \end{scope}

        \begin{scope}
            \draw (0,0) node [right=.25em,inner sep=0] (B)  {\includegraphics[width=3cm]{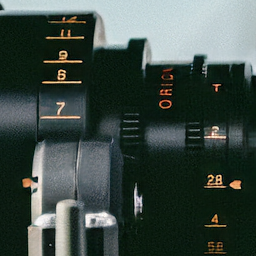}};
            \clip (B.south east) rectangle (B.north west);
            \draw (B)++(3,-2.1) node {\includegraphics[width=12cm]{fig/crop_clic400984b87394ada6d9627ed918908986_hificlo_numberchnaged}};
            \draw (B.north) node [below,white] {Reconstructed};
        \end{scope}

        \draw (A.north-|-5.5,0) node [below right] (O) {\textit{\textbf{HiFiC Lo}}};

        \draw (O.west|-0,0)++(0,4ex) node [right] {\scriptsize{compressed:}};
        \draw (O.west|-0,0)++(0,2ex) node [right] {\scriptsize0.09 bpp}; 
        \draw (O.west|-0,0)++(0,0ex) node [right] {\scriptsize2040$\times$1365};

        \draw (O.west|-0,0)++(0,-4ex) node [right] {\scriptsize{cropped scene:}};
        \draw (O.west|-0,0)++(0,-6ex) node [right] {\scriptsize{64$\times$64 {($0.15\%$)}}};

    \end{tikzpicture}

    \medskip

    \begin{tikzpicture}
        \begin{scope}
            \draw (0,0) node [left=.25em,inner sep=0] (A)  {\includegraphics[width=3cm]{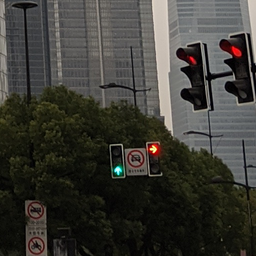}};
            \clip (A.south east) rectangle (A.north west);
            \draw (A)++(-.3,1.2) node {\includegraphics[width=12cm]{fig/crop_256_8f616c54067d0399619749dcad59e24b_original}};
            \draw (A.north) node [below,white] {Original};
        \end{scope}

        \begin{scope}
            \draw (0,0) node [right=.25em,inner sep=0] (B)  {\includegraphics[width=3cm]{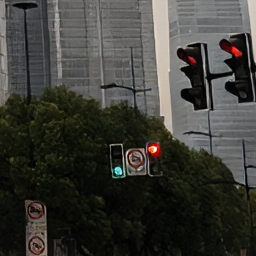}};
            \clip (B.south east) rectangle (B.north west);
            \draw (B)++(-.3,1.2) node {\includegraphics[width=12cm]{fig/crop_256_8f616c54067d0399619749dcad59e24b_param09}};
            \draw (B.north) node [below,white] {Reconstructed};
        \end{scope}

        \draw (A.north-|-5.5,0) node [below right] (O) {\textit{\textbf{CDC 0.9}}};

        \draw (O.west|-0,0)++(0,4ex) node [right] {\scriptsize{compressed:}};
        \draw (O.west|-0,0)++(0,2ex) node [right] {\scriptsize0.22 bpp};
        \draw (O.west|-0,0)++(0,0ex) node [right] {\scriptsize1152$\times$1920};

        \draw (O.west|-0,0)++(0,-4ex) node [right] {\scriptsize{cropped scene:}};
        \draw (O.west|-0,0)++(0,-6ex) node [right] {\scriptsize{64$\times$64 {($0.19\%$)}}};

    \end{tikzpicture}
    \caption{Category \catshape{}: Reconstructions contain changed contours.}
    \label{fig:ex-shape}
    \vspace{-2ex}
\end{figure}

\paragraph{\catshape} refers to reconstructions that differ in shape, potentially caused by biases in the retrieval augmentation process.
Semantic changes occur when the shape of an object conveys a semantic meaning, as illustrated in Fig.~\ref{fig:ex-shape}.
The top row shows a cropped image of a camera lens.
The change in shape causes the number $8$ in the original image to be mistaken as the number $6$ in the reconstruction.
In the bottom row, the shape of direction-specific traffic lights changes in the reconstruction from arrows to round lights.
This results in a change of semantic meaning, as a round green light typically indicates that drivers can proceed in any direction.

Although we have observed several instances of altered textures, we do not consider them as a distinct type for now because the alterations did not align with our definition of miscompressions, \ie{} changes in semantic meaning.
However, texture changes have been reported in the super-resolution literature~\cite[p.~25789]{li2024sed}, and we retain it as a potential extension to our taxonomy.

\begin{figure}
    \begin{tikzpicture}
        \begin{scope}
            \draw (0,0) node [left=.25em,inner sep=0] (A)  {\includegraphics[width=3cm]{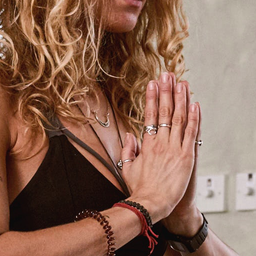}};
            \clip (A.south east) rectangle (A.north west);
            \draw (A)++(.1,.5) node {\includegraphics[width=6cm]{fig/clicacda8851def5fe87ebfe17f4ca90e55e_orig}};
            \draw (A.north) node [below,white] {Original};
        \end{scope}

        \begin{scope}
            \draw (0,0) node [right=.25em,inner sep=0] (B)  {\includegraphics[width=3cm]{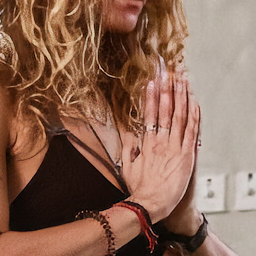}};
            \clip (B.south east) rectangle (B.north west);
            \draw (B)++(.1,.5) node {\includegraphics[width=6cm]{fig/clicacda8851def5fe87ebfe17f4ca90e55e_tattoodissapears.png}};
            \draw (B.north) node [below,white] {Reconstructed};
        \end{scope}

        \draw (A.north-|-5.5,0) node [below right] (O) {\textit{\textbf{HiFiC Lo}}};

        \draw (O.west|-0,0)++(0,4ex) node [right] {\scriptsize{original image:}};
        \draw (O.west|-0,0)++(0,2ex) node [right] {\scriptsize0.15 bpp};
        \draw (O.west|-0,0)++(0,0ex) node [right] {\scriptsize1228$\times$1840};

        \draw (O.west|-0,0)++(0,-4ex) node [right] {\scriptsize{cropped scene:}};
        \draw (O.west|-0,0)++(0,-6ex) node [right] {\scriptsize{128$\times$128 {($0.73\%$)}}};

    \end{tikzpicture}
    \caption{Miscompression of \emph{symbols} such as body adornments or religious jewelry increase the risk of semantic misinterpretation.}
    \label{fig:ex-jewlery}
    \vspace{-2ex}
\end{figure}

To classify the potential semantic impact of miscompressions across all categories, we define the \textsc{Symbol} modifier.
The consequence of miscompressions is elevated when the affected objects portray \emph{symbols}, \ie{} signs that carry specific meaning to human observers within a given social and cultural context~\cite{frutiger1989signs}.
Examples of symbols include letters, numerals, and signs, as well gestures, body adornments (\eg{} religious jewelry or clothing, wedding rings, etc.), traffic signs and lights, watch hands, logos, tattoos, graffiti, etc.
This list is not exhaustive, and identifying a \textsc{Symbol} is subjective and can be challenging, especially without knowledge of the cultural and societal context of the captured scene.
When symbols are involved, small changes in amplitude, geometry, or shape can completely alter the semantic meaning.
For instance, the miscompression of a plant in the bottom row of Fig.~\ref{fig:ex-geometry} is likely harmless, whereas missing jewelry, as shown in Fig.~\ref{fig:ex-jewlery} could lead to discord.

While screening our data, we have made a number of noteworthy qualitative observations.
First, we found that miscompressions commonly occur in small image details (see Fig.~\ref{fig:ex-geometry}), and a single image can contain multiple instances of miscompressions.
Notably, not every image contains miscompressions.
Images that depict single large objects against flat backgrounds are less susceptible.
Moreover, we find that \textit{CDC} is more likely to visibly destroy text, which reduces the risk of misinterpretation of incorrectly reconstructed text.
In general, miscompressions occur seemingly unpredictably, and are difficult to distinguish from authentic image details.

\section{\label{sec:discussion}Discussion}

In this section, we briefly reflect on the decisions that have shaped our taxonomy, then outline how it can be applied, before discussing the wider implications of miscompressions and closing with selected avenues for future research.

The definition of miscompressions based on textual descriptions is naturally subjective.
It depends on the observer with their experience and on the language, which defines concepts based on culture.
Intersubjectivity can be improved by asking multiple observers~\cite{smaling1992varieties}.
The language dependency aligns the definition with relevance in the given cultural context.

The proposed taxonomy of miscompressions is a first qualitative step towards mitigations.
The next step is to apply the taxonomy for the annotation of a larger dataset.
This will pave the way towards quantifying the prevalence of miscompressions and identifying influencing factors with statistical methods.
Importantly, such a dataset could be used to train models that detect and classify miscompressions, removing the human from the loop and allowing even greater scale.

Also image-to-text models can be useful tools to this end.
We presented it with a $256\times 256$ crop (bigger than in Fig.~\ref{fig:ex-amplitude}) that included the lower part of the tower and roofs of houses.
ChatGPT's description changed from \emph{``The image depicts a church steeple with a cross at the top, situated in a mountainous area. [\dots]''}
for the original image to \emph{``The image appears to be of a church steeple or a spire set against a mountainous backdrop. [\dots]''}
But this approach did not work for all of our examples.
For instance, there was no difference in the description of the watch in Fig.~\ref{fig:ex-amplitude}.
Using targeted object recognition methods to identify specific cases of miscompressions appears more promising, \eg{} the use of optical character recognition to identify miscompressed text.

Automatic detectors can be implemented as a safety net at encoding time to catch potential miscompressions.
This would allow an increase in the number of bits allocated to areas of the image where miscompressions loom.
The next step would be to use annotations of miscompressions as part of the training loss metric in order to harden future neural compression schemes against miscompressions.
Our taxonomy can be used to tailor these metrics to the types of miscompressions that should be avoided in a particular application, for example humans in the surveillance of public places and license plates in traffic surveillance.
Conversely, neural compression could be tuned to \emph{deliberately cause} miscompressions of, say, human faces as an integrated privacy-enhancing technology~\cite{lander2001evaluating, ilia2015face, li2017effectiveness}.

While research should strive to avoid miscompressions entirely, in the meantime it is crucial to deal with the existing risk in practice.
It is imperative to acknowledge the existence of miscompressions and explain the associated risks of misunderstandings and false accusations to end users of the technology.
Neural compression is not ready for use in safety and security critical applications, such as public surveillance or autonomous driving.
The benefit of bandwidth savings is disproportionate to the risk of wrongful convictions and potentially fatal accidents.
Worryingly, surveillance and autonomous driving are mentioned prominently in the motivation for the upcoming JPEG AI neural compression standard~\cite[p.~104]{JPEGAIstandard2023ascenso}.
In less critical applications, the use of neural compression should be documented.
Suitable annotation could be stored in image metadata, where professionals, such as photo journalists and forensic investigators, can find and interpret them.
A quantitative perceptual metric of miscompressions, similar to the metric for photo retouching~\cite{kee2011perceptual}, could be used in image captions, visible watermarks, or icons, and inform consumers about the potential presence of miscompressions.
Reliable methods to detect neural compression are needed to enforce such policies~\cite{bergmann2023frequency,bergmann2024forensic}.
Interesting open research problems remain:
Is it possible to detect instances of miscompressions in reconstructed images without access to the original?
Can we develop forensic methods to distinguish uncontrolled miscompressions from malicious manipulations?

\section{\label{sec:conclusion}Conclusion}
To our knowledge, this is the first study that compares multiple neural compression schemes for their susceptibility to produce semantically different reconstructions.
We raise awareness of this novel problem in forensics, propose a provisional taxonomy of what we call \emph{miscompressions}, and support it with existential evidence.
We hope that as this taxonomy develops, it will enable quantitative studies of automated detection, prevalence, influencing factors, and mitigations.

\bibliographystyle{IEEEbib}
\bibliography{HB_ARXIV_miscompressions.bib}

\begin{thebibliography}{10}

\bibitem{ABCNews_2016}
``{Boston Bombing Day 2},'' Apr 2016,
\newblock
  \url{https://abcnews.go.com/US/boston-bombing-day-improbable-story-authorities-found-bombers/story?id=38375726},
  last accessed: July 2024.

\bibitem{justicegov_2024}
``{Dzhokhar A. Tsarnaev},'' Mar 2024,
\newblock \url{https://www.justice.gov/usao-ma/tsarnaev-exhibits-day-2}
  exh\_29.pdf, last accessed: July 2024.

\bibitem{yang2023introduction}
Y.~Yang, S.~Mandt, and L.~Theis,
\newblock ``An introduction to neural data compression,''
\newblock {\em Foundations and Trends{\textregistered} in Computer Graphics and
  Vision}, pp. 113--200, 2023.

\bibitem{mentzer2020high}
F.~Mentzer, G.~Toderici, M.~Tschannen, and E.~Agustsson,
\newblock ``High-fidelity generative image compression,''
\newblock {\em NeurIPS}, 2020.

\bibitem{yang2024lossy}
R.~Yang and S.~Mandt,
\newblock ``Lossy image compression with conditional diffusion models,''
\newblock {\em NeurIPS}, 2024.

\bibitem{Agustsson_2017DIV2K}
E.~Agustsson and R.~Timofte,
\newblock ``{NTIRE} 2017 challenge on single image super-resolution: Dataset
  and study,''
\newblock in {\em CVPR}, 2017.

\bibitem{balle2018variational}
J.~Ball{\'e}, D.~Minnen, S.~Singh, S.~Hwang, and N.~Johnston,
\newblock ``Variational image compression with a scale hyperprior,''
\newblock in {\em ICLR}, 2018.

\bibitem{wallace1991jpeg}
G.~Wallace,
\newblock ``The {JPEG} still picture compression standard,''
\newblock {\em Communications of the ACM}, pp. 30--44, 1991.

\bibitem{lewis2009towards}
A.~Lewis and M.~Kuhn,
\newblock ``Towards copy-evident {JPEG} images,''
\newblock {\em GI Jahrestagung}, pp. 1582--1591, 2009.

\bibitem{shannon1948mathematical}
C.~Shannon,
\newblock ``A mathematical theory of communication,''
\newblock {\em The Bell System Technical Journal}, pp. 379--423, 1948.

\bibitem{ahmed1974DCT}
N.~Ahmed, T.~Natarajan, and K.R. Rao,
\newblock ``Discrete cosine transform,''
\newblock {\em IEEE Transactions on Computers}, pp. 90--93, 1974.

\bibitem{huffman1952method}
D.~Huffman,
\newblock ``A method for the construction of minimum-redundancy codes,''
\newblock {\em IRE}, pp. 1098--1101, 1952.

\bibitem{duan2022opening}
Z.~Duan, M.~Lu, Z.~Ma, and F.~Zhu,
\newblock ``Opening the black box of learned image coders,''
\newblock in {\em PCS}. IEEE, 2022, pp. 73--77.

\bibitem{ding2020image}
K.~Ding, K.~Ma, S.~Wang, and E.~Simoncelli,
\newblock ``Image quality assessment: Unifying structure and texture
  similarity,''
\newblock {\em TPAMI}, pp. 2567--2581, 2020.

\bibitem{NEURIPS2020_00482b9b}
S.~Bhardwaj, I.~Fischer, J.~Ball\'{e}, and T.~Chinen,
\newblock ``An unsupervised information-theoretic perceptual quality metric,''
\newblock in {\em NeurIPS}, 2020, pp. 13--24.

\bibitem{theis2017lossy}
L.~Theis, W.~Shi, Q.~Cunningham, and F.~Husz{\'a}r,
\newblock ``Lossy image compression with compressive autoencoders,''
\newblock {\em ICLR}, 2017.

\bibitem{balle2016end}
J.~Ball{\'e}, V.~Laparra, and E.~Simoncelli,
\newblock ``End-to-end optimized image compression,''
\newblock {\em arXiv preprint arXiv:1611.01704}, 2016.

\bibitem{goodfellow2014generative}
I.~Goodfellow, J.~Pouget-Abadie, M.~Mirza, B.~Xu, D.~Warde-Farley, S.~Ozair,
  A.~Courville, and Y.~Bengio,
\newblock ``Generative adversarial nets,''
\newblock {\em NeurIPS}, 2014.

\bibitem{sohl2015deep}
J.~Sohl-Dickstein, E.~Weiss, N.~Maheswaranathan, and S.~Ganguli,
\newblock ``Deep unsupervised learning using nonequilibrium thermodynamics,''
\newblock in {\em ICML}. PMLR, 2015, pp. 2256--2265.

\bibitem{ho2020denoising}
J.~Ho, A.~Jain, and P.~Abbeel,
\newblock ``Denoising diffusion probabilistic models,''
\newblock {\em NeurIPS}, pp. 6840--6851, 2020.

\bibitem{berthet2022ai}
A.~Berthet and J.~Dugelay,
\newblock ``{AI}-based compression: A new unintended counter attack on
  {JPEG}-related image forensic detectors?,''
\newblock in {\em ICIP}. IEEE, 2022, pp. 3426--3430.

\bibitem{berthet2023impact}
Alexandre Berthet, Chiara Galdi, and Jean-Luc Dugelay,
\newblock ``On the impact of {AI}-based compression on deep learning-based
  source social network identification,''
\newblock in {\em MMSP}. IEEE, 2023, pp. 1--6.

\bibitem{bergmann2023frequency}
S.~Bergmann, D.~Moussa, F.~Brand, A.~Kaup, and C.~Riess,
\newblock ``Frequency-domain analysis of traces for the detection of {AI}-based
  compression,''
\newblock in {\em IWBF}. IEEE, 2023, pp. 1--6.

\bibitem{bergmann2024forensic}
S.~Bergmann, D.~Moussa, F.~Brand, A.~Kaup, and C.~Riess,
\newblock ``Forensic analysis of {{AI}}-compression traces in spatial and
  frequency domain,''
\newblock {\em Pattern Recognit. Lett.}, pp. 41--47, 2024.

\bibitem{chen2023towards}
T.~Chen and Z.~Ma,
\newblock ``Towards robust neural image compression: Adversarial attack and
  model finetuning,''
\newblock {\em TCSVT}, 2023.

\bibitem{jalilian2022}
E.~Jalilian, H.~Hofbauer, and A.~Uhl,
\newblock ``Iris image compression using deep convolutional neural networks,''
\newblock {\em Sensors}, vol. 22, no. 7, 2022.

\bibitem{marr2010vision}
D.~Marr,
\newblock {\em Vision: A computational investigation into the human
  representation and processing of visual information},
\newblock MIT press, 2010.

\bibitem{hudelot2005symbol}
C.~Hudelot, N.~Maillot, and M.~Thonnat,
\newblock ``Symbol grounding for semantic image interpretation: From image data
  to semantics,''
\newblock in {\em ICCV}. IEEE, 2005, pp. 1875--1875.

\bibitem{minnen2020channel}
D.~Minnen and S.~Singh,
\newblock ``Channel-wise autoregressive entropy models for learned image
  compression,''
\newblock in {\em ICIP}. IEEE, 2020, pp. 3339--3343.

\bibitem{CLIC2020}
G.~Toderici, W.~Shi, R.~Timofte, L.~Theis, J.~Balle, E.~Agustsson, N.~Johnston,
  and F.~Mentzer,
\newblock ``Workshop and challenge on learned image compression (clic2020),''
\newblock in {\em CVPR}, 2020.

\bibitem{franzen1999kodak}
R.~Franzen,
\newblock ``Kodak photocd dataset,'' Nov 1999.

\bibitem{piva2013overview}
A.~Piva,
\newblock ``An overview on image forensics,''
\newblock {\em International Scholarly Research Notices}, p. 496701, 2013.

\bibitem{li2024sed}
B.~Li, X.~Li, H.~Zhu, Y.~Jin, R.~Feng, Z.~Zhang, and Z.~Chen,
\newblock ``{SeD}: Semantic-aware discriminator for image super-resolution,''
\newblock in {\em IEEE/CVF CVPR}, 2024, pp. 25784--25795.

\bibitem{frutiger1989signs}
A.~Frutiger,
\newblock {\em Signs and Symbols},
\newblock Weiss Verlag, 1989.

\bibitem{smaling1992varieties}
A.~Smaling,
\newblock ``Varieties of methodological intersubjectivity - the relations with
  qualitative and quantitative research, and with objectivity,''
\newblock {\em Quality and Quantity}, pp. 169--180, 1992.

\bibitem{lander2001evaluating}
K.~Lander, V.~Bruce, and H.~Hill,
\newblock ``Evaluating the effectiveness of pixelation and blurring on masking
  the identity of familiar faces,''
\newblock {\em JARMAC}, pp. 101--116, 2001.

\bibitem{ilia2015face}
P.~Ilia, I.~Polakis, E.~Athanasopoulos, F.~Maggi, and S.~Ioannidis,
\newblock ``Face/off: Preventing privacy leakage from photos in social
  networks,''
\newblock in {\em CCS}. 2015, pp. 781--792, ACM.

\bibitem{li2017effectiveness}
Y.~Li, N.~Vishwamitra, B.~Knijnenburg, H.~Hu, and K.~Caine,
\newblock ``Effectiveness and users' experience of obfuscation as a
  privacy-enhancing technology for sharing photos,''
\newblock {\em PACM Human-Computer Interaction}, pp. 1--24, 2017.

\bibitem{JPEGAIstandard2023ascenso}
J.~Ascenso, E.~Alshina, and T.~Ebrahimi,
\newblock ``The {JPEG} {AI} standard: providing efficient human and machine
  visual data consumption,''
\newblock {\em IEEE MultiMedia}, pp. 100--111, 2023.

\bibitem{kee2011perceptual}
E.~Kee and H.~Farid,
\newblock ``A perceptual metric for photo retouching,''
\newblock {\em PNAS}, pp. 19907--19912, 2011.

\end{thebibliography}

\end{document}